\documentclass[aps,prl,twocolumn,superscriptaddress]{revtex4-1}

\usepackage{graphicx}

\begin{document}


\title{Entropy production in a non-Markovian environment}



\author{Aki Kutvonen}
\affiliation{COMP Center of Excellence, Department of Applied Physics,
Aalto University School of Science, P.O. Box 11000, FI-00076 Aalto, Espoo, Finland}
\author{Tapio Ala-Nissila}
\affiliation{COMP Center of Excellence, Department of Applied Physics,
Aalto University School of Science, P.O. Box 11000, FI-00076 Aalto, Espoo, Finland}
\affiliation{Department of Physics, Brown University, Providence RI 02912-1843, U.S.A.}
\author{Jukka Pekola}
\affiliation{Low Temperature Laboratory (OVLL) and Department of Applied Physics, Aalto University Scool of Science, P.O. Box 13500, FI-00076 Aalto, Espoo, Finland}



\date{May 26, 2015}

\begin{abstract}
Stochastic thermodynamics and the associated fluctuation relations provide the means to extend the fundamental laws of thermodynamics to 
small scales and systems out of equilibrium. The fluctuating thermodynamic variables are usually treated in the context of either
isolated Hamiltonian evolution, or Markovian dynamics in open systems. However, there is no reason a priori why the Markovian
approximation should be valid in driven systems under non-equilibrium conditions. In this work we introduce an explicitly 
non-Markovian model of dynamics of an open system, where the correlations between the system and the 
environment drive a subset of the environment out of equilibrium. Such a an environment gives rise to a new type of non-Markovian entropy production
term. Such non-Markovian components must be taken into account in order to recover the fluctuation relations for entropy. As a concrete example, 
we explicitly derive such modified fluctuation relations
for the case of an overheated single electron box.
\end{abstract}


\maketitle


The fundamental laws of thermodynamics are among the most powerful postulates of physics due to their universality. 
Motivated by recent developments in nanotechnology, 
a great effort has been made to extend thermodynamics to microscopically small systems and to 
non-equilibrium processes \cite{Jarzynski2011,Seifert2012,Bustamante2005,Collin2005}. 
These efforts are commonly formulated in the form of stochastic thermodynamics, which culminates into fluctuation relations connecting
extensive thermodynamic variables such as work, free energy and entropy \cite{Seifert2008, Seifert2005,Jarzynski1997,Sagawa2010,Crooks1999}.

Stochastic thermodynamics and the associated fluctuation relations can be formulated through trajectory dependent, fluctuating quantities, 
typically assuming Markovian evolution for open systems interacting with an ideal heat bath. The Markovian assumption means that all the degrees of freedom
of the infinite environment relax to the equilibrium state at time scales much faster than the those of the system, and thus there are no memory effects caused by the coupling
of the system and the environment. However, since by definition of an open system there is always a coupling term in the Hamiltonian of the total (isolated) system that
contains both background and system degrees of freedom, Markovian evolution is strictly speaking an approximation. 
In small systems there is no reason {\it a priori} that there is a division to clearly separate fast background
and slow system degrees of freedom,
and thus non-Markovian effects like memory of dynamics may become important.

Some recent works have examined entropy production under non-Markovian dynamics by 
considering different types of memory or by using other thermodynamic arguments \cite{Hasegawa2011,Garcia2012,Campisi2011,Speck2007}. 
However, the crucial assumption behind these models is that the environment cannot change during the process \cite{Speck2007}. 
In this work we consider a non-Markovian system, in which some degrees of freedom in the environment are 
influenced by the interaction with the system, and thus the environment must be explicitly taken into account. 
We extend the classical stochastic thermodynamics approach to include the non-Markovian dynamics arising from such effects. 

In particular, we show how violation of the Markovian approximation of the environment leads to generation of additional
entropy. Our approach allows the identification and separation of the total entropy production into standard Markovian and new non-Markovian components. The
existence of a non-Markovian component of entropy production leads to modifications of the standard fluctuation relations for entropy. As a concrete example of this,
we consider the single electron box (SEB), which has proven to be an excellent test bench for studying thermodynamics at small scales \cite{Koski2013,Koski2014,Koski2014b,Saira2012,Pekola2013}. 
In particular, we demonstrate that an overheated SEB is an experimentally feasible system with well defined non-Markovian dynamics, and derive new fluctuation relations for entropy production
in such systems.

We follow the standard approach of stochastic thermodynamics (see e.g. \cite{Esposito2010,Seifert2008} and references therein) 
and consider a system whose degrees of freedom evolve under the influence of a time dependent control parameter 
$\lambda(t)$ from $x_0=x(0)$ to $x_f=x(t_f)$ along a trajectory $x(t)$. The time reverse processes are given by $\tilde{x}(t)=x(t_f-t)$ and $\tilde{\lambda}(t)=\lambda(t_f-t)$. 
In stochastic thermodynamics the total entropy generation of any given trajectory $x(t)$ is defined as
\begin{equation}
\Delta S_T=\Delta S_S+\Delta S_F = \ln{ \frac{p_I[x_0]}{p_F[\tilde{x}_0]}}+ \ln{ \frac{p[x(t)|x_0]} {\tilde{p}[\tilde{x}(t)|\tilde{x}_0]}},
\label{eq:Stotmat}
\end{equation}
where $\Delta S_S$ is the change in the system entropy and $\Delta S_F$ is the entropy flow to the environment (or medium entropy generation), 
$p[x(t)|x_0]$ and $\tilde{p}[\tilde{x}(t)|\tilde{x}_0]$ are the probabilities of the forward and reverse path, 
and $p_I[x_0]$ and $p_F[\tilde{x}_0]$ are the probabilities of the initial and final states, respectively. From the conservation of probability, it immediately follows that
\begin{equation}
\langle e^{-\Delta S_T} \rangle =1.
\label{eq:Stotint}
\end{equation}

If the dynamics are stochastic and governed by a Markovian master equation, the entropy flow $\Delta S_F$ can be expressed as:
\begin{equation}
\Delta S_F[x]= \ln{\prod_{j=1}^N \frac{W_{x_j,x_{j-1}} (\lambda_j)}{W_{x_{j-1},x_{j}} (\lambda_j)}},
\label{eqmark}
\end{equation}
where $W_{x_j,x_{j-1}} (\lambda_j)$ is the transition rate from $x_{j-1}=x(t_{j-1})$ to $x_{j}$ at control parameter value $\lambda_j=\lambda(t_j)$. 
Furthermore, if the rates are connected by the local detailed balance condition (LDB)
\begin{equation}
\frac{W_{x_j,x_{j-1}} (\lambda_j)}{W_{x_{j-1},x_{j}} (\lambda_j)}=e^{\beta Q_j},
\label{eq:LDB}
\end{equation}
where $Q_j$ is the heat dissipation from the system to the bath at $t=t_j$ and $\beta=1/T_B$ is the inverse temperature of the heat bath, the entropy flow becomes
\begin{equation}
\Delta S_F[x]=\Delta S_F^m[x]=\beta  Q[x],
\label{eq:Sflowm}
\end{equation}
where $Q=\sum_j Q_j$ is the heat flow from the system during trajectory $x$. We note that if the initial distribution $p_I$ is that of a canonical equilibrium state, 
the total entropy is given by the dissipated work $\Delta S_T=W[x]-\Delta F$, and Eq. (\ref{eq:Stotmat}) becomes the Crooks fluctuation relation and Eq. (\ref{eq:Stotint}) the Jarzynski equality \cite{Crooks1999,Jarzynski1997}.

The Markovian approximation, where the rate matrix is only a function of $\lambda$ (cf. Eq.  (\ref{eqmark})) and the LDB condition of Eq. (\ref{eq:Sflowm}), 
together require that the environment is an ideal heat bath with a relaxation time $\tau_{B}$ much shorter than any other time scale in the total system. 
However, since there is always a finite interaction between the system and the environment, there are always some correlations between their dynamics.
In small systems, such correlations may become important if there is no clear separation of time scales into a fast, infinitely large environment and a slow, finite system. We now consider a setup beyond the Markovian assumption, namely the case where the environment as a whole is not an ideal heat bath described by equilibrium temperature $T_B$. We assume that there exists a subset of degrees of freedom of the environment which are correlated with the dynamics of the system and do not relax to equilibrium defined by $T_B$ on a time scale faster than that of the system. We call these degrees of freedom the \emph{non-equilibrium subsystem} (NE) of the environment 
which cannot be described by the bath's thermal equilibrium distribution during the drive and must therefore be explicitly taken into account.

In general, a rigorous study of such non-equilibrium fluctuations in the environment requires a full knowledge of the 
time evolution of all the degrees of freedom inside the NE. However, if it 
has an internal relaxation time $\tau_{NE}$ much shorter than the characteristic relaxation time scale of the system $\tau_{S}$, we can simplify the problem. In addition, we assume that the NE is weakly coupled 
to the rest of the environment such that all the degrees of freedom in the NE mutually relax to a (non-equilibrium) state at time scales $\tau_{NE} \ll \tau_{S}$. We note that even if the relaxation inside the NE is the fastest time scale,  Markovian evolution does not follow here since the NE is finite and due to energy deposition from the system during the trajectory $x$, the NE will evolve through quasi-equilibrium states different from the equilibrium state of the rest of the environment. That is to say that the transition $x_i \to x_{i+1}$ is affected by the earlier transitions $\{x_j \to x_{j+1} \}$, where $j<i$.

We will further assume that the
NE contains sufficiently many degrees of freedom such that it can be described by a time dependent probability distribution
function (density operator) $\rho_{NE}(t)$ with an effective temperature $T(t)$. 
Furthermore, as the relaxation time $\tau_B$ is short and the correlation between the NE and 
the rest of the environment is weak, we assume that the environment
outside of the NE is described by the equilibrium distribution 
$\rho^{eq}$ and acts as a Markovian heat bath. In this case the total distribution $\rho_T(t)$ can be 
generally written as $\rho_T (t)=\rho_{S; NE} (t) \otimes \rho^{eq}$, where $\rho_{S; NE} (t)$ is the joint probability distribution
of the system and the NE. Further, since we assume that $\tau_{NE} \ll \tau_S$, we can separate this joint distribution function
into its two constituents and write the total distribution as
\begin{equation}
\rho_T (t)=\rho_S (t) \otimes \rho_{NE}(t) \otimes \rho^{eq}.
\label{eq:prodform}
\end{equation}
%
As a result the total trajectory entropy $\Delta S_T$ is no longer given by its 
Markovian component $\Delta S^m_T=\Delta S_F^m+\Delta S_S$. Instead, we have additional source of entropy, $S_{NE}(t)=-\ln{\rho_{NE}}$, coming from the fact that in addition to $\rho_S$, the probability distribution $ \rho_{NE}(t)$ changes in time. 

Returning to the dynamics of the system, 
the time scale separation assumed here means that the system is effectively driven in an environment controlled by the NE at
an effective temperature $T(t)$ instead of the bath temperature $T_B$. Therefore, the transition rates no longer satisfy the LDB condition of Eq. (\ref{eq:LDB}), but instead they should satisfy the condition
\begin{equation}
\frac{W_{x_j,x_{j-1}} (\lambda_j)}{W_{x_{j-1},x_{j}} (\lambda_j)}
=e^{(\beta+\Delta \beta_j) Q_j},
\label{eq:effDB}
\end{equation}
where $\Delta \beta_j \equiv 1/T(t_j)-1/T_B$. As long as the total system is described by Eq. (\ref{eq:prodform}) between the transitions, 
the NE can be integrated out giving rise to transition rates for the system states in Eq. (\ref{eq:effDB}). However, for example, if the number of degrees of freedom in the NE is not large enough, $\Delta \beta_j$ may not be associated with any physical temperature, but should be defined through Eq. (\ref{eq:effDB}) as a
parameter modifying the LDB condition. 
Assuming Eq. (7) holds and integrating out the equilibrium part gives the associated entropy flow $\Delta S_F$ (Eq. (\ref{eqmark})) as
\begin{equation}
\Delta S_F=\Delta S_F^m+ \Delta S_F^{nm}
=\beta  Q+\Delta \beta Q,
\label{nmphysentr}
\end{equation}
where now an additional {\it non-Markovian} contribution $\Delta S_F^{nm}$ appears in addition to the Markovian component $\Delta S_F^m= \beta  Q$, cf. Eq. (\ref{eq:Sflowm}). We note that even if the entropy flow can always be mathematically written as in Eq. (\ref{nmphysentr}),  the physical 
interpretation of $\Delta \beta Q$ in the general case may not be clear.

Since the heat bath and the NE are coupled, there can also be heat flow $Q^B$ from the NE to the bath. The associated entropy component is given by
\begin{equation}
\Delta S_F^{B}=\beta  Q^B.
\end{equation}
%

\begin{figure}
    \includegraphics[width=0.5\textwidth]{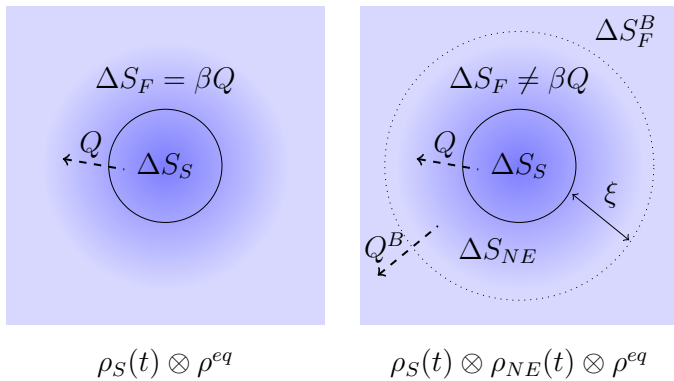}
    \caption{A schematic of the comparison between entropy generation in the two cases considered in the text. The left panel illustrates a purely Markovian system, in which the environment is assumed to stay in 
equilibrium during the process. The right panel shows the case of an NE, with 
a measure $\xi$ that now constitutes a subset of the bath degrees of freedom. The NE can be driven outside of equilibrium due to the coupling to the system leading to an additional entropy production (see text for details).
}
    \label{fig1}
\end{figure}

Adding the non-Markovian sources of entropy production to $\Delta S_T$, we can conclude that the total entropy production in now given by
\begin{equation}
\Delta S_T=\Delta S_S+\Delta S_F^m +\Delta S_F^{nm}+ \Delta S_{NE}+\Delta S_F^B,
\label{eq:totentsplit}
\end{equation}
where the sum of the first two terms is the Markovian component of the entropy production $\Delta S_T^{m}$, and the sum of the last three terms is the 
non-Markovian counterpart $\Delta S_T^{nm}$. The different sources of entropy production are schematically illustrated in Fig. \ref{fig1}.

To make the arguments presented above concrete, we next proceed to consider a physically realistic example, the single electron 
box (SEB) \cite{Averin1986}, where single-electron transitions are externally induced between two metallic islands. 
In recent experiments various fluctuation relations have been measured on a SEB in \cite{Saira2012,Koski2013}, and it has been harnessed for experiments to convert information into energy (Maxwell's demon) in Refs. \cite{Koski2014,Koski2014b}. The SEB is particularly suitable for studies of non-equilibrium statistical phenomena due to the limited number of relevant degrees of freedom, clear separation of different time scales, and well defined and easily measured energy scales.

\begin{figure}
    \includegraphics[width=0.3\textwidth]{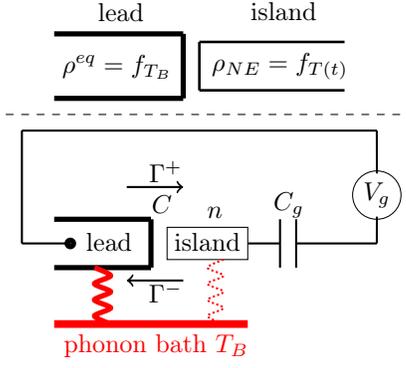}
    \caption{Schematic of a single electron box. The number of excess electrons on the island $n$, is a stochastic variable governed by the tunnelling rates to ($\Gamma^+$) and from ($\Gamma^-$) the island, which depend on the control parameter $V_G$, the gate voltage. As the island has a finite heat capacity, and since it is weakly coupled to the substrate, the electrons on the island tend to occupy a 
    Fermi distribution $f_T(t)$ at an effective temperature $T(t)$ which differs from the bath temperature. The other side of the tunnel junction is a reservoir with essentially infinite heat capacity and strong coupling to the phonon bath, and thus the electrons in this lead are Fermi distributed at the equilibrium temperature $T_B$.}
    \label{fig2}
\end{figure}

In our case the SEB is composed of a metallic island coupled to a metallic electrode (lead) by a tunnel junction and to another electrode by a capacitor, see Fig. \ref{fig2}. The relevant part of the Hamiltonian of the system is that of its Coulomb energy, given by $H_S(n,n_g) = E_C (n-n_g)^2$, where $E_C=e^2/(2C_\Sigma)$, $e$ is the electron charge, and $C_\Sigma=C+C_g+C_0$ is the total capacitance 
which is given by the sum of those of the tunnel junction $C$, the gate capacitance $C_g$, and the self capacitance $C_0$ of the island. 
The constant $E_C$ gives the unit of energy to charge the box by an extra electron. The control parameter $n_g=C_g V_g/e$ is given by the scaled gate voltage $V_g$ driving the electrons between the islands. The stochastic variable, $n$, is the number of extra electrons on the island of the box and should be considered as the system variable. Note, however, that in a metallic SEB the total number of electrons is large, of the order of $10^9$, whereas $n$ describes the deviation from the particular equilibrium number on the island. In a well-controlled experiment $n$ can assume only two values, say $n=0$ and $n=1$.  To achieve this regime, the SEB needs to be operated at low enough temperatures, $k_BT < E_C$, and the values of the control parameter $n_g$ vary within one period (amplitude $< 1$) only. A detailed discussion on the energetics and the different thermodynamic variables in the SEB can be found in Refs. \cite{Pekola2012,Averin2011}.

In the present case we consider the SEB in an experimentally feasible situation, where the metallic island is small and thus the heat capacity of the island is finite \cite{Pekola2013}. The collective of $10^9$ electrons on the island forms the NE of the setup which can be driven outside of equilibrium due to the small size of the metallic island. The relaxation of non-equilibrium excitations created by energy deposition of a tunnelling electron (electron-electron relaxation rate) on both sides of the tunnel junction is the fastest time scale in the problem ($\tau_{NE}$). In experiments the internal relaxation time $\tau_{NE}$ is of order of $10^{-9}$ s while the the relaxation time back to the equilibrium by the coupling to the phonon bath of the substrate on which the SEB is fabricated is of order of $10^{-4}$ s. Therefore, the electrons on the small metallic island of the tunnel junction adopt a Fermi distribution at a different effective temperature during the drive. We note that this temperature is a result of the stochastic energy deposition from tunnelling electrons on the small island due to the drive by the control gate. The state of the whole system is thus described by a product form of type $\rho_T(t)=\rho_S^n(t) \otimes f_{T(t)}(E_I) \otimes f_{T_B}(E)$ (cf. Eq. (\ref{eq:prodform})), where $\rho_S^n(t)$ is the distribution of excess electrons $n$ on the island at time $t$, $f_{T(t)}(E_I)$ is the Fermi distribution of electron energy $E_I$ on the island at an 
effective temperature $T(t)$, which depends on the energies of the previous tunnelling events.  Therefore the process is non-Markovian.
Finally,  $f_{T_B}(E)$ is the Fermi distribution on the lead side of the tunnel junction at the equilibrium temperature $T_B$. 
Thus the SEB setup here is a concrete, physical realisation of a well-defined NE, as discussed in the general theoretical setting. 
More details on the operation of an SEB in the this regime can be found in Refs. \cite{Saira2010,Pekola2013}.

We next proceed to study dissipation in a tunnelling process. Standard Fermi golden rule calculations yield the tunneling rates
\begin{eqnarray}
\Gamma^\pm(t)_{T(t)}=\int_{-\infty}^{\infty}\gamma^\pm_{T(t)} (E,t) dE,
\label{tunnelingrates}
\end{eqnarray}
where $+$ denotes tunneling in, $-$ tunnelling out, and $\gamma^\pm_{T(t)} (E,t)dE$ is the differential electron tunneling rate within the energy interval $dE$ around $E$ (from the Fermi level of the lead) such that:
\begin{equation}
\gamma^\pm_{T(t)} (E,t)=\frac{1}{e^2R_T}f_{T_B}(\pm E)\{1-f_{T(t)}(\pm [E + \Delta U(t)]),
\end{equation}
where $R_T$ is the tunneling resistance and
$\Delta U$ is the chemical potential difference across the junction. In equilibrium the rates $\Gamma^\pm$ are LDB connected, but if $T(t) \ne T_B$, the LDB condition is broken and condition of type of Eq. (\ref{eq:effDB}) holds.

In a tunneling event the total dissipated heat is given by
\begin{equation}
Q_{tot}^\pm=Q_I^\pm+Q_L^\pm=\pm \Delta U,
\end{equation}
where the heat deposited to the lead in a tunneling event is given by $Q_L^\pm= \mp E$ and the heat deposited to the island is $Q_I^\pm=\pm(E+\Delta U)$. Equivalently, the total dissipated heat in a tunnelling (in) event is the energy released by the system $\Delta U=H_S(0,n_g)-H_S(1,n_g)=E_C(2n_g-1)$.

Since in a tunnelling event an energy $Q_I^\pm$ is deposited to local temperature $T(t)$ and energy $Q_L^\pm$ to temperature $T_B$, we can write the entropy flow in a tunneling event as
\begin{equation}
S_F^\pm=(\beta+\Delta \beta)Q_I^\pm+\beta Q_L^\pm,
\label{eq:SF}
\end{equation}
where the non-Markovian component is given by $S_F^{nm,\pm}=\Delta \beta Q_I^\pm$, $\beta=1/k_B T_B$ and $\Delta \beta \equiv 1/k_B T - 1/k_B T_B$. Another contribution to the non-Markovian sources is the entropy of the NE, i.e. the electron population of the island, given by $S_{I}=-\ln{ f_{T}}$, where $f_T$ is the probability distribution of degrees of freedom inside the NE ($\rho_{NE}$), the Fermi function of the island at temperature $T$. 
In a tunneling event at time $t$, an electron with energy $E$, which changes the temperature of the island from $T_i$ to $T_{i+1}$, induces an entropy change of
\begin{equation}
S_I^\pm
=\log{\frac{f_{T_i}[\pm(E+\Delta U (t))]}{f_{T_{i+1}}[\pm(E+\Delta U (t))]}}.
\label{eq:SI}
\end{equation}
The total change of $S_F$ and $S_I$ in a trajectory are given by the sum over individual tunneling events (Eqs. (\ref{eq:SF}) and (\ref{eq:SI})), which we denote by $\Delta S_F$ and $\Delta S_I$, respectively.

The standard trajectory system entropy change is given by
\begin{equation}
\Delta S_S=\log{\frac{p_I[n_I]}{p_F[n_F]}},
\label{eq:SS}
\end{equation}
where $p_I(n_I)$ and $p_F(n_F)$ are the probabilities to sample the initial and final states $n_I$ and $n_F$, respectively.

Initially, the NE is coupled to the bath, and the initial temperature $T_I$ is sampled from distribution $p_I(T_I)$ which is in equilibrium normally distributed with a variance $k_B T^2/C$, where $C$ is the heat capacity of the island. Thus, in addition to the tunneling events, $S_I$ can change due to the heat transfer from the bath. The associated entropy is given by
\begin{equation}
\Delta S_I^T=\ln{\frac{p_I^T(T_I)}{p_F^T(T_F)}},
\label{eq:SIT}
\end{equation}
where $p_F(T_F)$ is the probability to sample the final temperature $T_F$. In typical experiments, the heat capacity of the small metallic island is of the order of $1000 k_{B}$, and the operating temperature around $100$ mK, resulting to initial temperature fluctuations of the order of a few mK. Depending on the drive protocol, the speed of the drive in particular, the heat dissipation per tunneling electron is of the order of  $k_{B}$ resulting to temperature fluctuations of the order of $0.1$ mK per tunneling event. We note that the electron-phonon coupling can be neglected at short time scales of the process such that we can neglect the term $\Delta S_F^B$ in Eq. (\ref{eq:totentsplit}). 

Proceeding to the dynamics of the SEB, the probability that there exist no tunneling in ($+$) or tunneling out ($-$) events from $t_i$ to $t_{i+1}$, while the temperature of the island is $T$, is given by
\begin{equation}
P_{NT}^\pm [t_i,t_{i+1},T]=e^{-\int_{t_i}^{t_{i+1}} \Gamma^\pm_{T}(t')dt'}.
\end{equation}
The probability that an electron with energy $E$ at time $t$ tunnels is given by $\gamma^\pm_{T} (E,t)$. 
We denote the direction of tunneling event $i$ in a realization of $n$ tunneling events as $i_n \in \{+,-\}$. 
Because of the two possible states, $(i+1)_n=-i_n$. Furthermore, the time labeling of temperature $T(t)$ is chosen such that $T_I=T(t_0)$, and 
at tunneling $i$ at time $t_i$ the temperature is $T(t_{i-1})$. Tunneling event at $t_i$, where a heat $Q_I=\pm(E-\mu_I)$ is deposited to the island, changes the temperature of the island to $T(t_i)=T(t_{i-1})+Q_I/C$. Thus the probability of a realization of a path $X$ is given by
\begin{eqnarray}
P(X)   \equiv p_I[n_I] p_I^T(T_I)
P_{NT}^{1_n} [0,t_1,T_I] \times \nonumber \\
 \prod_{i=1}^{i=n}
\gamma_{T(t_{i-1})}^{i_n} (E_i,t_i)
P_{NT}^{(i+1)_n} [t_i,t_{i+1},T(t_i)],
\label{fpath}
\end{eqnarray}
where $t_{n+1}=t_f$ is the final time of the process and $t_{0}=0$ is the initial time. The sum of probabilities of all possible paths is normalised, i.e.,
\begin{equation}
\int P(X) dX \equiv
\sum_{n_I} \sum_{\{i_n\}}
 \prod_{i=0}^n \int_{t_i}^{t_{i+1}} \int_{-\infty}^{\infty} P(X) dt_i dE_i =1.
 \label{px}
\end{equation}

\begin{figure}
\includegraphics[width=0.4 \textwidth]{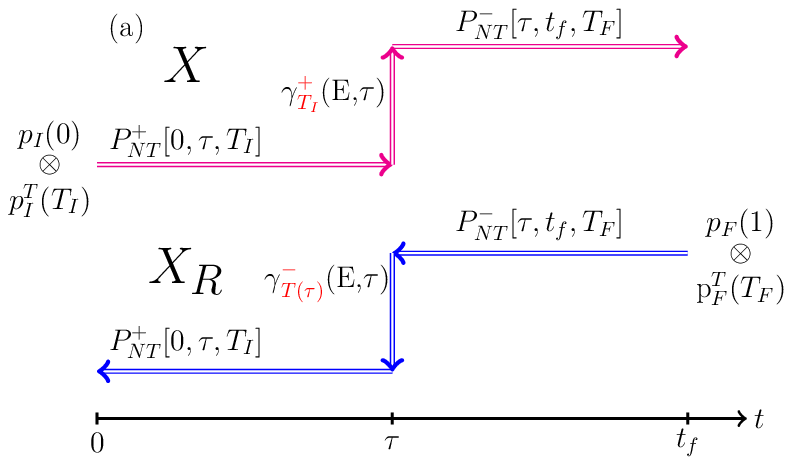}
\includegraphics[width=0.35 \textwidth]{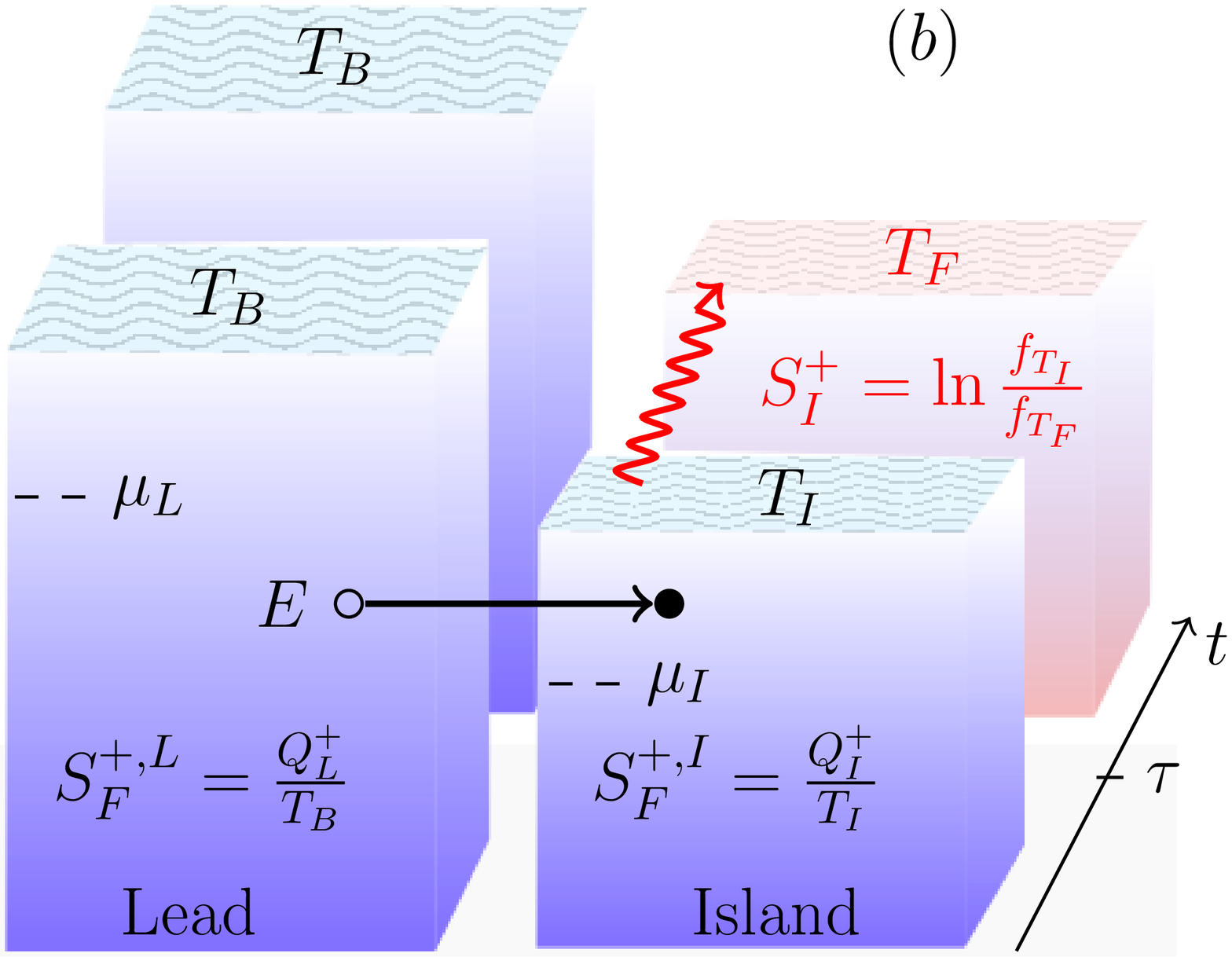}
\caption{(a) An example of a trajectory $X$ starting from state $n=0, T(0)=T_I$ followed by single tunneling-in event at time $t$ with energy $E$. Due to the relaxation of energy $Q^+_I=\mu_I-E$ the effective temperature on the island changes to $T_F$. The different terms denote the constituents of the total path probability. Trajectory $X_R$ demonstrates the reverse of $X$. The ratio of path probabilities yields $P(X)/P_R(X_R)=\exp{[\Delta S_T[X]]}$. 
(b) Entropy generation in a tunneling event, where heats $Q_L^+=\mu_L-E$ and $Q_I+=E-\mu_I$ are deposited to the lead and the island, respectively. The entropy flow in the event is given by $S_F^+=S_F^{+,L}+S_F^{+,I}=(\beta+\Delta \beta) Q^{+}_{tot}$. After the tunneling event at time $\tau$, the non-equilibrium excitation created by the tunnelling electron relaxes. As a result the electron energy distribution on the island changes from $f_{T_I}$ to $f_{T_F}$. The associated entropy change, the change of the NE entropy, is given by $S_I^+=\ln[{f_{T_I}(E+\Delta U)}/{f_{T_F}(E+\Delta U)}]$.}
\label{fig3}
\end{figure}

The probability of a time-reversed trajectory $P_R(X_R)$ is the probability to observe the time reversed trajectory $X_R$ of trajectory $X$ 
under time reversal of the external control $\lambda(t) \rightarrow \lambda(t_f-t)$. The trajectory $X_R$ starts from the final state of trajectory $X$. 
In addition, all the tunneling events are reversed. Thus we can write
\begin{eqnarray}
P_R(X_R) \equiv p_F[n_F] p_F^T[T_F]
P_{NT}^{(n+1)_n} [t_n,t_f,T_F] \times \nonumber \\
\prod_{i=1}^{i=n}
\gamma_{T(t_{i})}^{-i_n} (E_i,t_i)
P_{NT}^{i_n} [t_{i-1},t_{i},T(t_{i-1})].
\label{rpath}
\end{eqnarray}
The probability $P_R(X_R)$ is also normalized to unity. An example of $P(X)$ and the corresponding $P_R(X_R)$ is illustrated in Fig. \ref{fig3}.

Using the relation between the forward and reverse paths, we show in the Appendix that the associated probabilities satisfy the detailed fluctuation relation
\begin{equation}
\frac{P(X)}{P_R(X_R)}
=e^{\Delta S_T}
=e^{\Delta S_S+\Delta S_I^T+\Delta S_I+\Delta S_F}.
\label{eq:pathratio}
\end{equation}
This is our first main result, which allows to determine the total entropy production from the forward and reversed path probabilities. Furthermore, as we expect the total entropy production to satisfy the detailed fluctuation relation, Eq. (\ref{eq:pathratio}) supports the the claim that the non-Markovian sources of entropy production are given in a way we assumed. Using Eq. (\ref{eq:pathratio}) we obtain the integral fluctuation relation
\begin{equation}
\langle e^{-\Delta S_T} \rangle
= \int e^{-\Delta S_T}  P(X) dX= \int P_R(X_R) dX_R=1.
\label{expstot}
\end{equation}
 Furthermore, using the fluctuation relation for the Markovian entropy production $\Delta S_T^m=\Delta S_S+\Delta S_F^m$ yields
\begin{widetext}
\begin{equation}
\langle e^{-\Delta S_T^m} \rangle
=\langle e^{-\Delta S_T+\Delta S_T^{nm}} \rangle
= \int e^{-\Delta S_T[X]+\Delta S_T^{nm}[X]}  P(X) dX= \int e^{-\Delta S_T^{nm}[X_R]}P_R(X_R) dX_R=
\langle e^{-\Delta S_T^{nm}} \rangle_R,
\label{mentrexp}
\end{equation}
\end{widetext}
where we used the fact that entropy generation is odd under time reversal, $\Delta S_T^{nm}[X_R]=-\Delta S_T^{nm}[X]$. 
By $\langle \cdot \rangle _R$ we denote an average over the process where the control parameter (gate voltage) protocol is time reversed. 
In general the R.H.S. of Eq. (\ref{mentrexp}) is not equal to unity. Thus, the standard fluctuation relations which consider the entropy production only as a function of system degrees of freedom, are valid in absence of entropy production $\Delta S^{nm}_T$. If the correlations extend to the environmental degrees of freedom, the environment has to be measured as well to determine the total entropy production.

To summarize, we have argued here that for any open system interacting with its environment, the interaction between the system and the environment 
unavoidably creates non-Markovian correlations. If there is a clear time scale separation between the slow system and the fast degrees of freedom of an ideal (infinite) environment,
the Markovian approximation is acceptable. However, there are many systems where this may not be the case.
Here we have presented a quantitative way to take the
non-Markovian effects into account by assuming that part of the environmental degrees of freedom form a non-equilibrium subsystem which remains in a non-equilibrium state 
while transitions in the system take place, and the rest of the environment remains in its thermal equilibrium state. 
In this formulation, the non-equilibrium state of the environment causing the memory effects is explicitly taken into account allowing us to identify both Markovian and non-Markovian components of the total entropy production $\Delta S_T^m$ and $\Delta S_T^{nm}$, respectively. 
We have used the overheated SEB as an example, where this theoretical scenario is realized. The SEB allows a straightforward calculation and physical interpretation of the non-Markovian sources of entropy production. As our main results, we have derived detailed and integral fluctuation relations for the total entropy production within the SEB and showed how the Markovian and 
non-Markovian components of $\Delta S_T$ are related in terms of the forward and backward trajectories (Eq. (\ref{mentrexp})).

It is important to note that although our explicit calculations have been made for the SEB, 
the theoretical arguments in this article are generally valid for systems, where the state of the total system can be expressed as a product of the type 
of Eq. (\ref{eq:prodform}). Such systems include driven quantum mechanical systems, where the total system is under unitary evolution and the initial state is modelled by product state of the system and the environment density operators $\hat{\rho}_S(t) \otimes \hat{\rho}_E^{eq}$. In such systems the environment is driven as well and as a result we expect additional non-Markovian entropy production \cite{Aurell2014,Ankerhold2014,Schmidt2014}.
 
Acknowledgements: This research has been supported in part by the Academy of Finland through its Centres of 
Excellence Programs (project nos. 251748 and 250280).
We wish to thank Carlos Mejia-Monasterio, Elsi Laine, Samu Suomela and Takahiro Sagawa for useful comments.

\appendix

\section{Appendix: Derivation of Eq. (22) in the main text}

By implementing the definitions of forward and backward trajectories (Eqs. (19) and (21)) we obtain:
\begin{widetext}
\begin{equation}
\frac{P(X)}{P_R(X_R)}
=\frac{p_I[n_I] p_I^T(T_I)
P_{NT}^{1_n} [0,t_1,T_I]
\prod_{i=1}^{i=n} \gamma_{T(t_{i-1})}^{i_n} (E_i,t_i) P_{NT}^{(i+1)_n} [t_i,t_{i+1},T(t_i)]}
{p_F[n_F] p_F^T[T_F] P_{NT}^{(n+1)_n} [t_n,t_f,T_F] \prod_{i=1}^{i=n} \gamma_{T(t_{i})}^{-i_n} (E_i,t_i)
P_{NT}^{i_n} [t_{i-1},t_{i},T(t_{i-1})]}.
\end{equation}
The terms $P_{NT}$ cancel each other and the ratio of the terms $p_I$ and $p_F$ can be expressed with the use of Eqs. (16) and (17). Thus,
\begin{equation}
\frac{P(X)}{P_R(X_R)}
=e^{\Delta S_S+\Delta S_I^T} \prod_{i=1}^{i=n}  \frac{
\gamma_{T(t_{i-1})}^{i_n}(E_i,t_i)}
{\gamma_{T(t_{i})}^{-i_n} (E_i,t_i)}.
\label{eq0a}
\end{equation}
In order to simplify this expression, we first prove a local detailed balance type of equality
\begin{equation}
e^{- S_{F}^\pm(E,t) }  \gamma^\pm_{T}(E,t)= \gamma^\mp_{T}(E,t).
\label{eq0}
\end{equation}
By using the definition of $S_F^\pm$ (Eq. (14)), the L.H.S. of the equation above yields
\begin{eqnarray}
e^{- S_{F}^\pm(E,t) }  \gamma^\pm_{T}(E,t)
=\frac{1}{e^2R_T}e^{\pm \beta E}f_{T_B}(\pm E)e^{\mp (\beta + \Delta \beta)(E + \Delta U)}\{1-f_{T}[\pm (E + \Delta U(t)]\}.
\end{eqnarray}
By using the properties of Fermi functions, $1-f_{T}(E)=f_{T}(-E)$ and $e^{ (\beta + \Delta \beta)(E)}f_{T}(E)=f_{T}(-E)$ we obtain
\begin{equation}
\frac{1}{e^2R_T}e^{\pm \beta E}f_{T_B}(\pm E)e^{\mp (\beta + \Delta \beta)(E + \Delta U)}\{1-f_{T}[\pm (E + \Delta U(t)]\}
=\frac{1}{e^2R_T}f_{T_B}(\mp E)\{1-f_{T}[\mp (E + \Delta U(t)]\},
\end{equation}
which equals the R.H.S. of Eq. (\ref{eq0}). Another result we need is:

\begin{eqnarray}
e^{- S_{I}^\pm(t,E,T_1,T_2) } \gamma^\mp_{T_1}(E,t)
=\frac{f_{T_2}[\pm(E+\Delta U (t))]}{f_{T_1}[\pm(E+\Delta U (t))]} 
\frac{1}{e^2R_T}f_{T_B}(\mp E)\{1-f_{T_1}[\mp (E + \Delta U(t)]\} \nonumber \\
 =\frac{f_{T_2}[\pm(E+\Delta U (t))]}{f_{T_1}[\pm(E+\Delta U (t))]}
\frac{1}{e^2R_T}f_{T_B}(\mp E)\{f_{T_1}[\pm (E + \Delta U(t)]\} 
=\frac{1}{e^2R_T}f_{T_B}(\mp E)\{1-f_{T_2}[\mp (E + \Delta U(t)]\}
\label{tempchange}
\end{eqnarray}

where on the first line we used the definition of $S_I^\pm$ (Eq. (15)) and on the second line the Fermi function identity $1-f_{T}(E)=f_{T}(-E)$. RHS of Eq. (\ref{tempchange}) equals $\gamma^\mp_{T_2}(E,t)$, thus
\begin{equation}
e^{- S_{I}^\pm(t,E,T_1,T_2) } \gamma^\mp_{T_1}(E,t)
=\gamma^\mp_{T_2}(E,t).
\label{eq1}
\end{equation}
By combining Eqs. (\ref{eq0}) and (\ref{eq1}) we obtain
\begin{equation}
\frac{\gamma^\pm_{T_1}(E,t)}{\gamma^\mp_{T_2}(E,t)}
=e^{S_{F}^\pm+S_{I}^\pm}.
\label{localDB}
\end{equation}
By using Eqs. (\ref{eq0a}) and (\ref{localDB}) and the fact that the total entropy changes $\Delta S_F$ and $\Delta S_I$ are sums over the entropies 
$S_F^\pm$ and $S_I^\pm$ produced in single events, we arrive to the final result
\begin{equation}
\frac{P(X)}{P_R(X_R)}
=e^{\Delta S_S+\Delta S_I^T+\Delta S_I+\Delta S_F}
=e^{\Delta S_T}.
\end{equation}

\end{widetext}

\bibliography{nmseblib}

\end{document}